\RequirePackage{fix-cm}
\documentclass[twocolumn,epjc3]{svjour3}

\usepackage{changepage}
\usepackage{amsmath}

\usepackage{booktabs}
\usepackage{multirow}
\usepackage{lscape}
\usepackage{pdflscape}
\usepackage{url}
\usepackage{subfigure}
\usepackage{overpic}
\usepackage{color}
\usepackage{caption}
\usepackage{extdash}

\smartqed
\RequirePackage{graphicx}
\newcommand{\R}{\textsuperscript{\textregistered}}
\journalname{Eur. Phys. J. C}
\begin{document}

\title{Deep Machine Learning for the PANDA Software Trigger
}


\author{P.~Jiang\thanksref{e1,GSI,IMP,HFHF}
        \and
        K.~G\"otzen\thanksref{GSI} 
        \and
        R.~Kliemt\thanksref{GSI,RUB} 
        \and
        F.~Nerling\thanksref{GSI,HFHF,GUF} 
        and K.~Peters\thanksref{GSI,HFHF,GUF} 
\\        on behalf of the PANDA Collaboration
}

\thankstext{e1}{e-mail: p.jiang@gsi.de}


\institute{GSI Helmholtzzentrum f\"ur Schwerionenforschung GmbH Darmstadt, Germany\label{GSI}
  \and 
  Institute of Modern Physics, Chinese Academy of Sciences Lanzhou, China\label{IMP} 
  \and
  Helmholtz Forschungsakademie Hessen f\"ur FAIR (HFHF), GSI Helmholtzzentrum f\"ur Schwerionenforschung, Campus Frankfurt, Germany\label{HFHF}
  \and
  Ruhr-Universit\"at Bochum, Germany\label{RUB}
  \and
  Goethe-Universit\"at Frankfurt, Germany\label{GUF}
}

\date{Received: date / Accepted: date}

\maketitle

\begin{abstract}
Deep machine learning methods have been studied for the software trigger of the future PANDA experiment at FAIR, using Monte Carlo simulated data from the GEANT-based detector simulation framework PandaRoot. 
Ten physics channels that cover the main physics topics, including electromagnetic, exotic, charmonium, open charm, and baryonic reaction channels, have been investigated at four different anti-proton beam momenta.
Binary and multi-class classification together with seven different network architectures have been studied. 
Finally a residual convolutional neural network with four residual blocks in a binary classification scheme has been chosen due to its extendability, performance and stability. 
The presented study represents a feasibility study of a completely software-based trigger system.
Compared to a conventional selection method, the deep machine learning approach achieved a significant efficiency gain of up to 200\%, while keeping the background reduction factor at the required level of 1/1000. 
Furthermore, it is shown that the use of additional input variables can improve the data quality for subsequent analysis. This study shows that the PANDA software trigger can benefit greatly from the deep machine learning methods.

\end{abstract}

\section{Introduction}
\subsection{Motivation}

The antiProton ANnihilation at DArmstadt experiment (PANDA) at the Facility for Antiproton and Ion Research (FAIR)\ \cite{bib_FAIR} aims to explore hadron physics by means of antiproton induced reactions\ \cite{bib_PANDA,bib_PANDA2020}.
A wide array of physics topics and open questions is in the scope of PANDA, including strangeness physics, charm and exotics, nucleon structure and hadrons in nuclei\ \cite{PANDAPhysicsBook,PANDAPhaseOne}.
Some examples are the investigation of the charmonium spectrum, hyperon spectroscopy, electromagnetic form factors as well as the search for exotic states, e.g.\ hybrids, glueballs and hypernuclei.
At centre-of-mass energies between 2 and 5.5\ GeV the total $\bar{p}p$ cross section ranges between 50 and 100~mb\ \cite{Workman:2022ynf}, while the signals of interest have cross sections between a couple of microbarns and a few nanobarns or even picobarns.
Thus, with signal cross sections being many orders of magnitude smaller than the total proton-antiproton cross section, extracting the physics of interest is challenging in terms of background suppression.

With an expected average reaction rate of about 20~MHz, and with the average event size of 10-20~kB, the full data rate will be roughly 200~GB/s or more.
Because there is only a tiny fraction of events containing physics processes of interest, it would be inefficient to store all the data.
In order to identify the interesting events in an online environment, a sophisticated trigger system is necessary.
The PANDA triggering system is foreseen to reduce the rate of events to be stored by an approximate factor of 1000 to 20~kHz, reducing the stored data to 200~MB/s, which still leads to about 1~PB stored data per year.

Other experiments such as E835\ \cite{E760,E835}, ATLAS\ \cite{ATLAS}, BaBar\ \cite{Babar}, or BESIII\ \cite{BESIII} have different approaches to trigger and perform event selection.
Detector coincidence combined with clean background conditions, e.g. in $e^+e^-$ collisions, allow for more or less straightforward triggering decisions.
In PANDA, the kinematic similarity of signal and background reactions combined with a wide physics interest and the high interaction rate puts the key challenge on the trigger system and consequently the selections have to be 
made on full event candidates with high-level information.
Similar to experiments like LHCb\ \cite{LHCb,LHCb3}, ALICE\ \cite{ALICE_Trigger} or CMS\ \cite{CMS}, PANDA will use a Software Trigger system based on fully reconstructed event information.
To cope with the challenges, high-level properties need to be obtained from the reaction data for proper signal-background separation, and deep machine learning methods are studied for the PANDA Software Trigger to improve performance compared to a conventional cut-and-count method.

\subsection{Experiment Environment}
\begin{figure}[tp!]
    \centering
    \includegraphics[width=0.99\columnwidth]{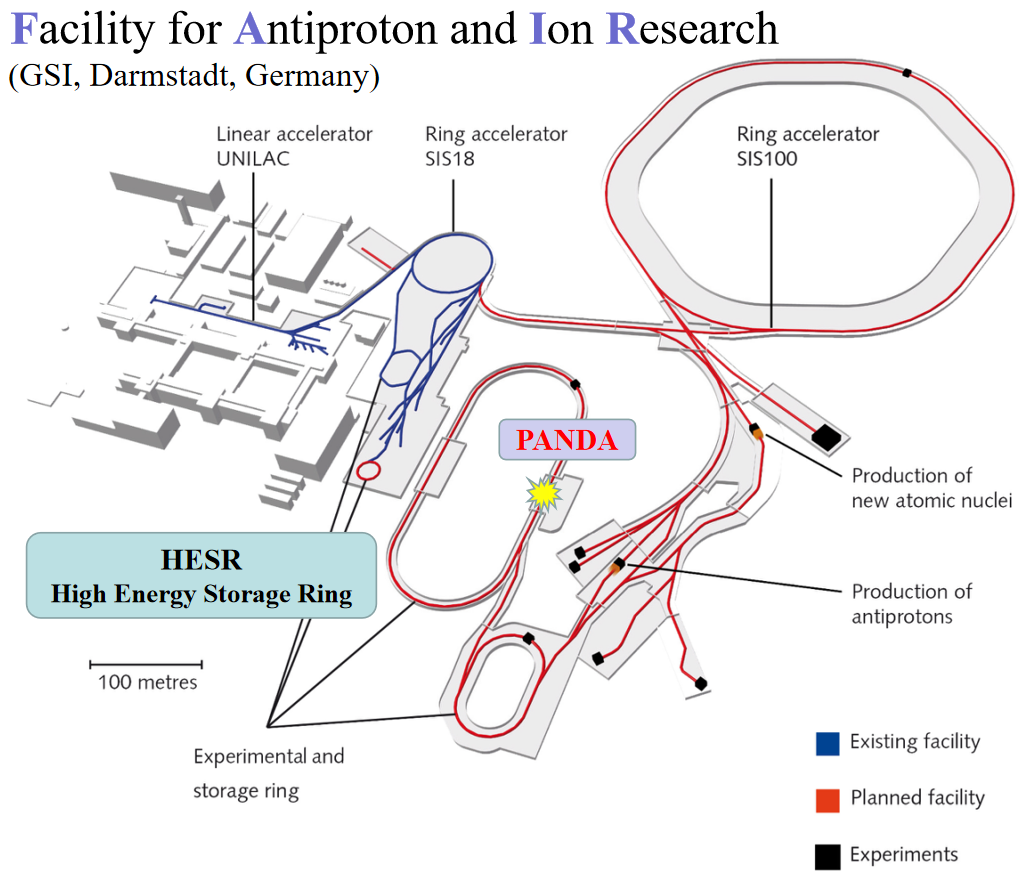}
    \hrule
    \includegraphics[width=1\columnwidth]{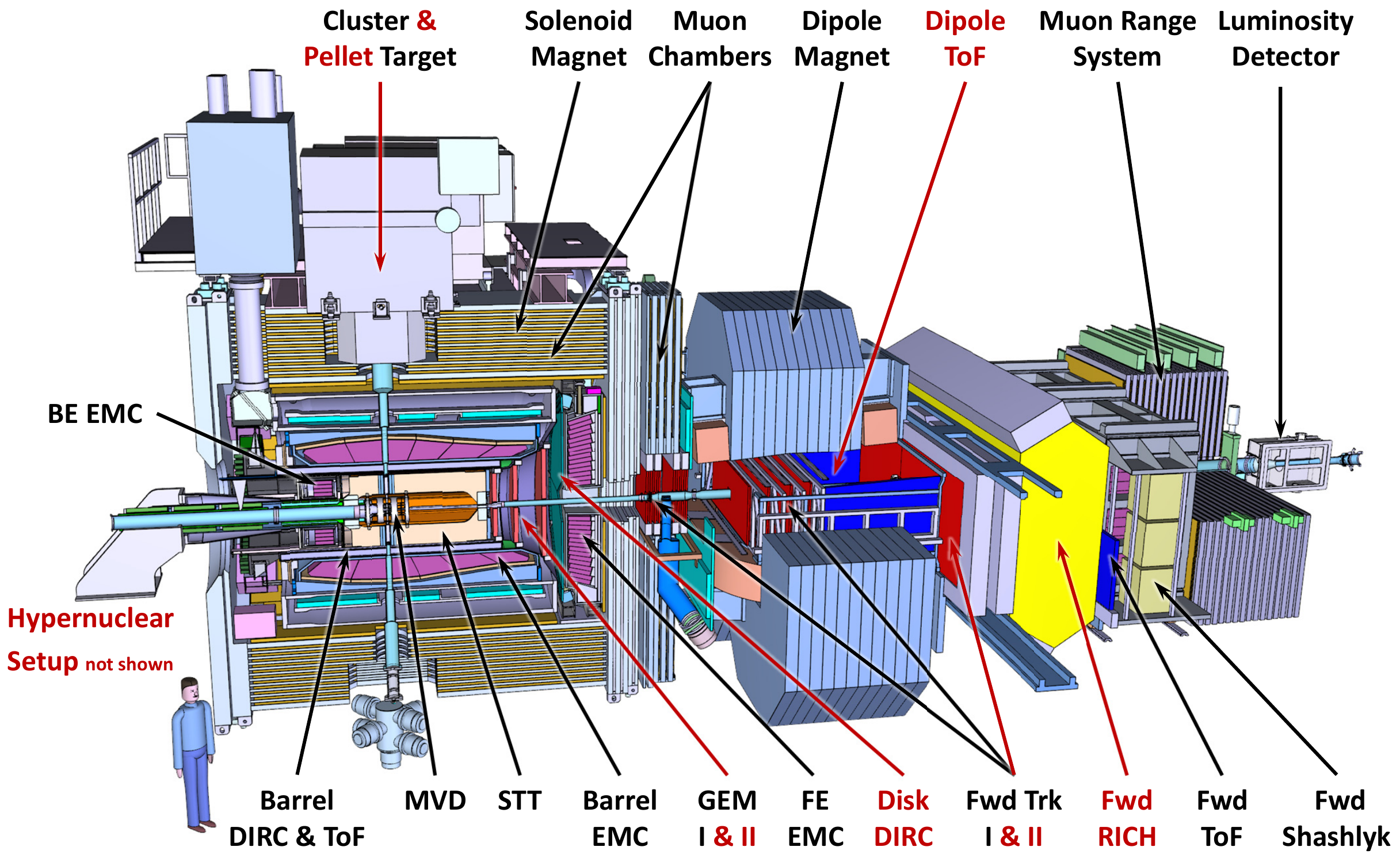}
    \caption{Overview of the FAIR facility (top). The PANDA experimental setup (bottom) with the initial detectors (in black) and staged upgrades (in red) \cite{bib_PANDA2020}(and references therein).}\label{Fig.FAIR}\label{Fig.PANDA}
\end{figure}

FAIR is an international accelerator facility being under construction in  Darmstadt, Germany\ \cite{bib_FAIR,bib_PANDA2020}.
It expands the existing accelerator complex in a large scale, shown in Figure~\ref{Fig.FAIR}.
An antiproton beam will be prepared in a cascade of accelerators, making use of almost the whole facility.
The synchrotron High-Energy Storage Ring (HESR)~\cite{bib_HESR} will store, cool and accelerate the antiprotons, which will have momenta in the range between 1.5 GeV/$c$ and 15 GeV/$c$, corresponding to centre-of-mass energies in the antiproton-proton reactions at PANDA between about 2 and 5.5 GeV.
With a luminosity of up to $L = 2 \cdot 10^{32} \:\rm{cm^{-2}s^{-1}}$ in the ``high luminosity mode'' and with a beam momentum resolution of $dp / p<2 \cdot 10^{-5}$ in the ``high resolution mode'' HESR is an essential component for PANDA that allows for precise measurements.

The PANDA experiment, shown in Figure~\ref{Fig.PANDA}, consists of
two spectrometers, one barrel-shaped surrounding the interaction point, the other covering the forward direction, both providing measurements for precise charged particle tracking and particle identification as well as electromagnetic calorimetry.

\subsection{Software Trigger in PANDA}

\begin{figure}[tp!]
    \centering
    \includegraphics[width=1\columnwidth]{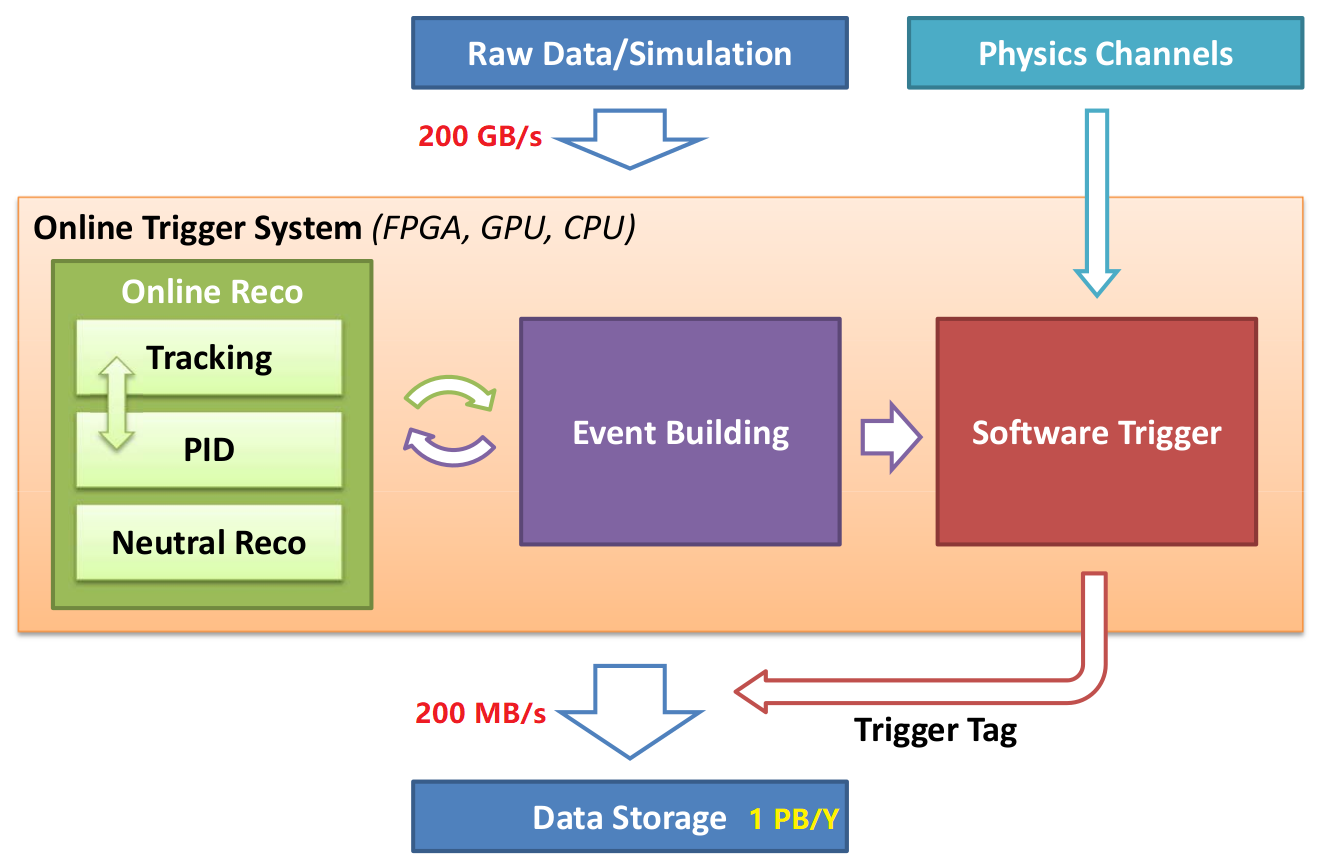}
    \caption{Schematic overview of the complete PANDA online trigger system.}\label{Fig.Schematic}
\end{figure}

The PANDA online trigger system will consist of online reconstruction, event building, and the Software Trigger, as illustrated in the schematic in Figure~\ref{Fig.Schematic}.
First, the online reconstruction of neutral particles and charged particles will be performed as completely as possible using fast algorithms, and the Particle IDentification (PID) information will be assigned to the corresponding tracks.
This will be combined with the event reconstruction process, possibly in an iterative manner, to provide the online event candidate with reconstructed final state particle information to the Software Trigger module.
The event candidate will then be processed in the Software Trigger module by the selection algorithms. 
Then the event candidates will be tagged when being consistent with a signal signature.
Finally, an event candidate will be written to the data storage if any of the trigger algorithms accepts it to be a signal event.

\section{Physics channels}
The PANDA experiment has a rich physics program, so that the composition of active trigger signatures will be adapted depending
on the current physics aim\ \cite{PANDAPhysicsBook,PANDAPhaseOne}.
Therefore, the PANDA Software Trigger system needs to identify many types of physics reactions and offer a high flexibility in configuration.
\begin{table}[bp!]
    \centering 
    \caption{List of ten physics channels for the Software Trigger study together with the corresponding codes used in this document and the trigger signatures.}  
    \scriptsize
    \begin{tabular}{p{0.22\columnwidth}|p{0.25\columnwidth}p{0.06\columnwidth}p{0.26\columnwidth}}
        \hline\noalign{\smallskip}
        Physics topic & Reaction channel & Code & Trigger\\
        \noalign{\smallskip}\hline\noalign{\smallskip}
        Electromagnetic & $\bar{p}p \rightarrow e^+e^-$ & $ee$ &  $\bar{p}p \rightarrow e^+e^-$  \\
        \noalign{\smallskip}\hline\noalign{\smallskip}
        Exotics         & $\bar{p}p \rightarrow \phi_{(1)}\phi_{(2)}$   & $Phi$ & $\phi \rightarrow  K^+K^-$ \\
        \noalign{\smallskip}\hline\noalign{\smallskip}
        Charmonium  & $\bar{p}p \rightarrow \eta_c\pi^+\pi^-$   & $Etac$ & $\eta_c \rightarrow K_S^0K^-\pi^+$ \\
        & $\bar{p}p \rightarrow J/\psi\,\pi^+\pi^-$  & $J2e$ & $ J/\psi \rightarrow e^+e^- $  \\
        & $\bar{p}p \rightarrow J/\psi\,\pi^+\pi^-$  & $J2mu$ & $ J/\psi \rightarrow \mu^+\mu^- $ \\
        \noalign{\smallskip}\hline\noalign{\smallskip}
        Open charm  & $\bar{p}p \rightarrow D^0\bar{D^0}$   & $D0$ & $D^0 \rightarrow K^-\pi^+$ \\
        & $\bar{p}p \rightarrow D^+D^-$   & $Dch$ & $D^+ \rightarrow K^-\pi^+\pi^+$ \\
        & $\bar{p}p \rightarrow D_s^+D_s^-$   & $Ds$ & $D_s^+ \rightarrow K^+K^-\pi^+$ \\

        \noalign{\smallskip}\hline\noalign{\smallskip}
        Baryons  & $\bar{p}p \rightarrow \Lambda \bar{\Lambda}$  & $Lam$ & $\Lambda \rightarrow p\pi^-$ \\
        & $\bar{p}p \rightarrow \Lambda_c \bar{\Lambda_c}$  & $Lamc$ & $\Lambda_c \rightarrow pK^-\pi^+$ \\

        \noalign{\smallskip}\hline\noalign{\smallskip}
        Background  & $\bar{p}p \hspace{0.05 in} generic  $ & DPM & - \\

        \noalign{\smallskip}\hline\noalign{\smallskip}
    \end{tabular}\label{tab.Channels}
\end{table}

For this study a total of ten channels are considered to verify the feasibility of the  Software Trigger inspired by the physics motivation given in the PANDA Physics Book\ \cite{PANDAPhysicsBook}.
These ten physics channels have simple and clear decay modes in their respective field, spanning a number of event topologies which may occur in the experimental runs, while covering the main physics topics, namely exotic hadrons, charmonium, open charm, and baryon states, including the resonances $\phi$, $\eta_c$, $J/\psi$, $D^0$, $D^+$, $D_s^+$, $\Lambda$, and $\Lambda_c$.
\begin{table}[bp!]
    \centering 
    \caption{Summarised list of input observables.}
\begin{tabular}{p{0.22\columnwidth} | p{0.65\columnwidth}}

\hline\noalign{\smallskip}
Category             & Description\\

\noalign{\smallskip}\hline\noalign{\smallskip}
                                      & - momentum $p$ of reconstructed candidate (lab/cms) \\
Candidate kinematics & - transverse momentum $p_t$ of reconstructed candidate \\
                                      & - polar angle $\theta$ of reconstructed candidate (lab/cms) \\

\noalign{\smallskip}\hline\noalign{\smallskip}
Daughter kinematics  & - kinematic variables from each daughter of the candidate \\
                                      & - Electron/Muon/Kaon/Pion/Proton PID probability of each daughter \\

\noalign{\smallskip}\hline\noalign{\smallskip}
                                      & - maximum (transverse) particle momentum in event  \\
Event info \& multiplicities          & - sum of (transverse) momenta of charged particles in event (cms) \\
                                      & - sum of cluster energies in EMC \\
                                      & - maximum of cluster energies in EMC \\
                                      & - Particle multiplicities \\
                                      & - Minimum/maximum/sum of momenta \\
                                      & - Number of loose ($P_{\rm PID} >~0.25$) candidates \\

\noalign{\smallskip}\hline\noalign{\smallskip}
                                      & - Aplanarity of the event\\
Event shape                           & - Sphericity of the event\\
                                      & - Magnitude of the event thrust vector\\
                                      & - Reduced Fox-Wolfram Moments 1-4 \\
\noalign{\smallskip}\hline
\end{tabular}\label{tab.Observables}
\end{table}
The physics topics, the reaction physics channels together with the corresponding codes and the trigger signatures are listed in Table~\ref{tab.Channels}. 
Signal events are generated with EvtGen\ \cite{EvtGen} using a simple phase space decay model (PHSP) as well as the VSS (decay of vector to two scalar particles) and VLL (decay of vector to two charged leptons) models for the decays of $\phi$ and $J/\psi$, respectively. 
More complex decay patterns e.g.\ in the Dalitz decay channels are omitted in order to populate the phase space more evenly for data quality studies.
Generic inelastic background reactions are generated by the Dual Parton Model (DPM)\ \cite{DPM}, which models the various production cross sections in antiproton-proton reactions.
These background events have to be rejected effectively by the triggering algorithm.

For each reaction type, a specific selection procedure, the trigger line, is defined.
To determine the performance, the individual trigger lines are under investigation as well as the complete trigger system for the ten reaction types to be tagged simultaneously.

A trigger line includes the reconstruction of a certain composite resonance / particle candidate and the classification whether this composite candidate originates from signal or background events.
A comprehensive list of input quantities (Table~\ref{tab.Observables}) is computed based on reconstructed properties, momenta, angles, particle identification probabilities of the composite candidates as well as final state particles involved.
Furthermore, event specific variables such as minimal and maximal momenta, momentum sums, event planarity and sphericity, thrust magnitude, Fox-Wolfram moments \cite{EventShape} and multiplicities are deduced, to further support the trigger decision.

One trigger line consists of four parts:
\begin{itemize}
    \item Reconstruction of composite candidates by final state combinatorics
    \item Preselection with a broad cut on the invariant mass of the candidate of $\pm10\sigma$ around its peak position
    \item Calculation of the necessary observables, Table~\ref{tab.Observables}
    \item Selection by algorithm (focus of this work)
\end{itemize}

The selection target is determined by achieving the required background suppression factor.
In this study, the overall required suppression factor is $s = 1/1000$, i.e.\ the number of background events is suppressed by 99.9\% with all active trigger lines acting simultaneously.
Applying $n_{\mathrm{trig}}$ trigger lines simultaneously, the background suppression for each trigger line is required to satisfy $s_{\mathrm{i}} = s / n_{\mathrm{trig}}$.
In turn, due to the possibility of different trigger lines accepting the same background event, the actual total background suppression factor $s_{\mathrm{all}}$ can be a little bit better, such that $s_{\mathrm{all}} \leq s = \sum s_{\mathrm{i}}$.

\section{Data preparation}
The data for the physics channels and background are obtained by Monte-Carlo simulations with the PandaRoot software\ \cite{PandaRoot}.
This simulation and reconstruction software framework 
is based on the publicly available software package FairRoot\ \cite{FairRoot} v18.2.0 and the the external package collection FairSoft in version ``jun19p1''.

EvtGen\ \cite{EvtGen} provides the particles for propagation through the detector volumes with the GEANT\ 3 transport modeller\ \cite{GEANT3}, which is followed by detailed detector simulations, digitisation and reconstruction.
Per particle candidate, tracking, calorimetry and PID information, such as particle energy loss, time of flight, Cherenkov angle, EMC energy deposit, is being provided.
All events are self-contained and no event time relevant effects, such as event mixing or incomplete events, are considered.
In Table~\ref{tab.Numbers}, the number of simulated events is shown for the chosen channels where the centre-of-mass energy permits the reaction.
Also the Monte-Carlo truth information, matched to the reconstructed particle candidates, is provided to distinguish combinatorial background from signal events for the training stage.
\begin{table*}[t]
  \centering  
  \caption{Numbers of simulated events of each physics channel at each energy, given in million events. The dashes mark the cases, where the reaction is energetically not possible.}
  \begin{tabular}{c|cccccccccccc}
    \toprule
    $\sqrt{s}[ GeV]$ &  $ee$ & $Phi$ & $Etac$ & $J2e$ & $J2mu$ & $D0$ & $Dch$ & $Ds$ & $Lam$ & $Lamc$ & DPM \\
    \midrule   
    2.4 & 3.1 & 1.3 & - & - & - & - & - & - & 2.4 & - &  20.0 \\
    3.8 & 2.4 & 1.2 & 6.0 & 2.7 & 1.2 & 1.2 & 1.4 & - & 1.3 & - & 20.0 \\
    4.5 & 2.2 & 1.2 & 6.6 & 2.3 & 1.2 & 1.2 & 1.3 & 1.6 & 1.2 & - & 20.0 \\
    5.5 & 2.2 & 1.2 & 7.7 & 2.1 & 1.2 & 1.2 & 1.2 & 1.3 & 1.2 & 1.7 & 20.0 \\
    \bottomrule 
  \end{tabular}\label{tab.Numbers}
\end{table*}

\section{Methods}
Our figure of merit for comparing different approaches to the triggering process will be the triggering efficiency, defined as the ratio of the number of triggered events ($N_{\rm{trig}}$) to the number of events passing the detector reconstruction, combinatorics and a 10$\,\sigma$ preselection mass window ($N_{\rm{rec}}$),
$$\epsilon = \frac{N_{\rm{trig}}}{N_{\rm{rec}}}$$
with $\sigma$ being the width/resolution of the individual reconstructed resonance.
The triggering efficiency can be defined for two cases. Individually the efficiency focuses on the selection performance of each trigger line, serving one particular physics channel.
This is a useful quantity to optimise the triggering algorithms in question.
For the complete trigger setup, the total efficiencies will be affected by cross-tagging. 
Here events can be rejected by their intended trigger line, but are accepted accidentally by one or more of the other trigger lines. This is an unintentional but fortunate effect.
The total triggering efficiency is the practically more relevant measure for the experiment.

Since we require a fixed background reduction for each neural network (trigger line) individually, the resultant signal trigger efficiencies already represent
a comparable measure for network performance. As an additional quality measure we provide the integral (AUC = area-under-curve) of the receiver-operator-characteris\-tics (ROC) 
being more common in Machine Learning context, while not being of large practical value for this study.  

\subsection{Cut-and-count method}
The conventional benchmark trigger scheme is following a cut-and-count approach employing an optimised set of trigger line specific one-dimensional cuts on the measured observable distributions. 

In order to identify this set of criteria all input observables are
evaluated individually for each trigger line. We integrate the distributions of both signal and background events for each of the $n$ observables in both directions up to
a threshold value retaining 90\% of the signal events. This threshold defines the cut to be applied. From these $2n$ possible cuts, we select and apply the one
leading to the largest suppression of background events. This procedure is iterated until the total background suppression matches the required
factor. The last criterion in the set is adapted in the way, that the background requirement is exactly met, so that the corresponding
signal efficiency can be larger than 90\%. 

The complete trigger configuration is given by 30 sets of one-dimensional selection criteria, one for each trigger line at each accessible centre-of-mass energy.

\subsection{Deep learning methods}
In order to let a neural network decide on which event belongs to one of the signal channels or to background, the data has to be presented in a sensible choice of observables. 
The same set of input variables (Table\ \ref{tab.Observables}) used for the cut-and-count approach is being used for direct comparison. Eventually this set of variables is extended by additional event shape observables to determine the optimal performance achievable using all available quantities.
The neural network forms its answer in a single number or a set of numbers in case of a multiple classification.
Performing a selection cut on this output will determine the rejection rate of background events as well as the efficiency of signal event acceptance.

The PyTorch framework\ \cite{pytorch} is chosen to provide the underlying functionality to build the neural networks, which are trained and evaluated with the prepared simulation data, roughly in a 1:1 ratio between the training and testing data sets.

\section{Neural network optimisation}
Neural network setups are quite diverse and mostly tailored for the problem at hand. For the Software Trigger, several choices are made based on the performance, especially the signal efficiency and background reduction.
Individual networks are optimised by:
\begin{itemize}
    \item Network depth\ \cite{DNN}
    \item Width or conventional kernel size of each layer
    \item Choice of the optimiser, learning rate and related items, e.g.\ momentum\ \cite{overviewGD}, 
    \item Choice of the activation function \cite{Sigmoid,ReLU,LeakyReLU,GeLU,SeLU,SoftPlus}, 
    \item Weight decay generalisation\ \cite{L2} 
    \item Weight initialisation\ \cite{Xavier}.
\end{itemize}

\subsection{Multi-class and binary approaches}
The first question investigated is if a single network serving all trigger lines simultaneously performs better than a single network for each physics channel.
Allowing multi-class classification would mean fewer but bigger networks to be trained. Binary classification in contrast allows for easier adaption and modification of the list of channels.
The multi-class approach reduces the number of required training experiments for each trigger setup to find the optimal network size at the cost of longer training times due to the increased complexity.

A set of Dense NNs\ \cite{DNN} (DNN) is used to study this issue exemplarily for one setting at $\sqrt s = 4.5\,\mathrm{GeV}$ by the means of a Bayesian approach for the best triggering efficiency\ \cite{Bayesian}.
The results are shown in Table~\ref{tab.multiclass_binary}, where the individual trigger efficiencies serve as figure of merit.
Here, only signal candidates that match the Monte-Carlo truth are under investigation, eliminating combinatorial effects.
In those channels where the triggering efficiency is not close to 100\%, the binary classification outperforms the multi-class approach by up to a factor of two.

Therefore, we choose the binary classification approach with one neural network per trigger line and energy point.

\begin{table}[bp!]
    \centering  
    \caption{The comparison on the individual trigger efficiencies of truth matched events between multi-class classification and binary classification based on a DNN.}
    \begin{tabular}{ c | c | c }
        \toprule
        Channels & Multi-class[\%] & Binary[\%] \\
        \midrule   
        $ee$ & 100.0 & - \\
        $Phi$ & 87.9 & - \\
        $Etac$ & 48.1 & 76.2 \\
        $J2e$ & 98.6 & - \\
        $J2mu$ & 99.9 & - \\
        $D0$ & 69.9 & 93.4 \\
        $Dch$ & 48.4 & 78.6 \\
        $Ds$ & 32.7 & 73.7 \\
        $Lam$ & 48.1 & 94.7 \\
        \bottomrule   
    \end{tabular}\label{tab.multiclass_binary}   
\end{table}

\subsection{NN type selection}
In order to identify the optimal network architecture and meta-configuration, seven different types of networks are studied,
listed in Table~\ref{tab.NNs}: A dense neural network (DNN),  a convolutional neural network (CNN)\ \cite{CNN}, both with and without residual blocks\ \cite{ResNet,ResNeXt,shatteringGradient}, a CNN with bottleneck residual blocks\ \cite{1x1} as well as 1D and a 2D Long-Short-Term-Memory network\ \cite{LSTM}.

\begin{figure*}[p]
    \begin{minipage}[t]{0.5\linewidth}
        \centering
        \begin{overpic}[width=\linewidth,tics=10,trim=0 0 0 20,clip]
        		{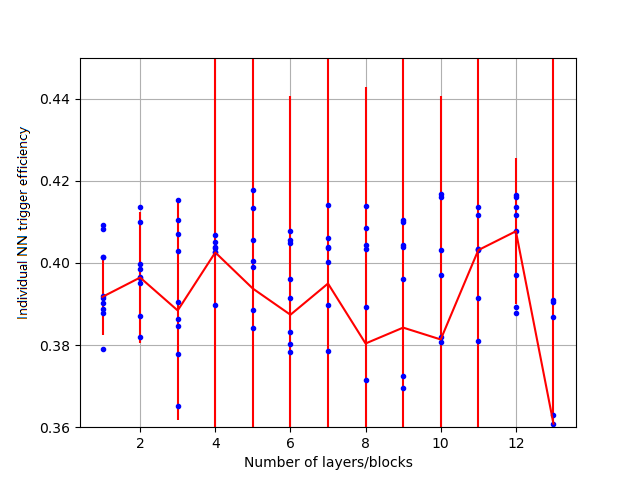}
          \put(15,60) {\fcolorbox{black}{white}{DNN}}
        \end{overpic}\\
        \begin{overpic}[width=\linewidth,tics=10,trim=0 0 0 20,clip]
        		{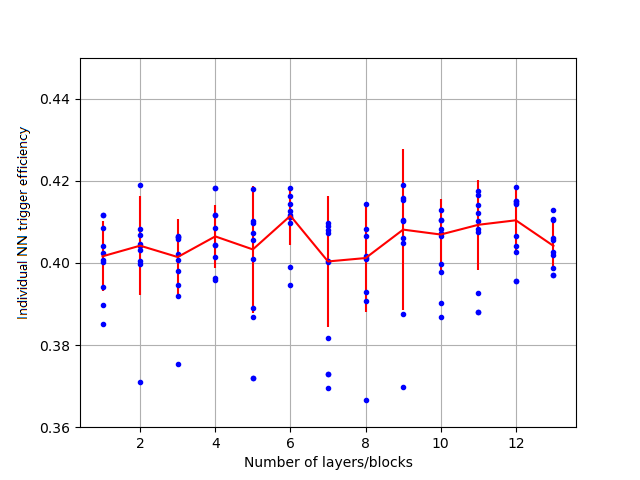}
          \put(15,60) {\fcolorbox{black}{white}{CNNRes}}
        \end{overpic}
    \end{minipage}
    \begin{minipage}[t]{0.5\linewidth}
        \centering
        \begin{overpic}[width=\linewidth,tics=10,trim=0 0 0 20,clip]
        		{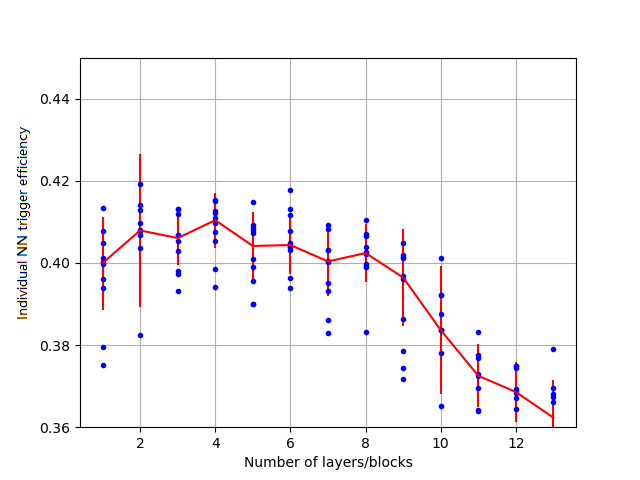}
          \put(15,60) {\fcolorbox{black}{white}{CNN}}
        \end{overpic}\\
        \begin{overpic}[width=\linewidth,tics=10,trim=0 0 0 20,clip]
        		{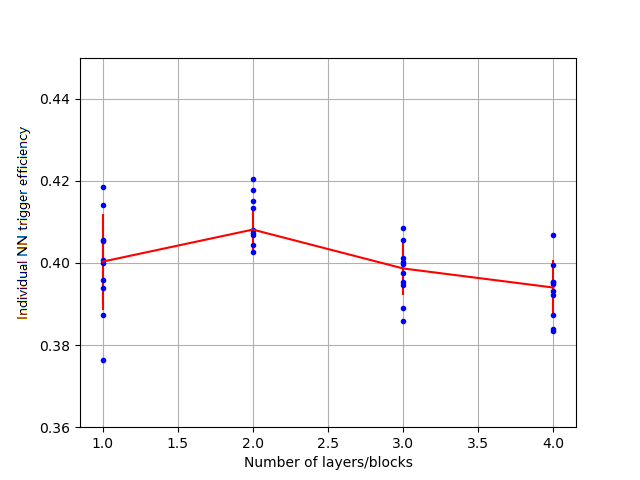}
          \put(15,60) {\fcolorbox{black}{white}{LSTM1D}}
        \end{overpic}
    \end{minipage}
    \caption{Individual trigger efficiencies (blue dots) for the channel $Dch$ at $5.5$~GeV/$c$ for four of the seven NN types with varying depth parameters as well as 
    the median with standard deviation (red makers and bars). }\label{Fig.numBlocks4Compare}
    \centering
    \includegraphics[width=2\columnwidth]{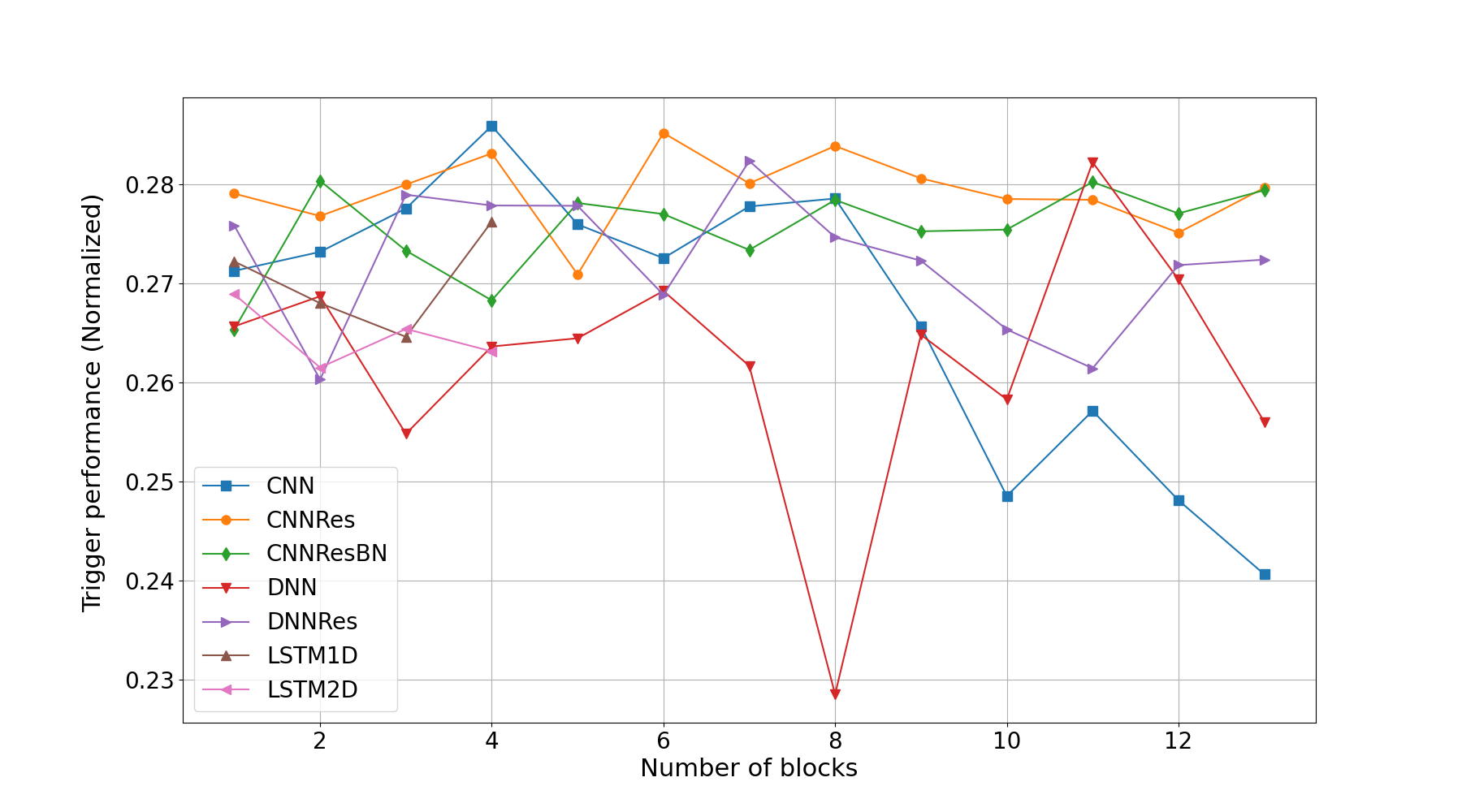}
    \caption{Normalised triggering performance $\hat\epsilon$ as function of the used layers/blocks in the seven network types, combined for the three channels $Etac$, $Dch$ and $Lamc$ at $5.5$~GeV/$c$.}\label{Fig.NNcomparison}
\end{figure*}

Some of the trigger lines perform reasonably well and show stable results under all kind of network configurations due to particularities of the decay kinematics. 
For example a $J/\psi$ decaying into two leptons will leave two strongly correlated high-momentum tracks in the detector, which is significantly different to the average multi-pion background event and in consequence always
leads to high selection quality.

For a meaningful optimisation, these ``simple'' cases are ignored, and
the networks are optimised in depth and layer size for a high triggering efficiency for the three more challenging 
channels $Etac$, $Dch$ and $Lamc$.
For instance, the depth is optimised by obtaining the trigger performance as function of used layers/blocks, and ten runs are carried out for each NN framework of a certain depth.
The performance is evaluated by the individual efficiencies 
as well as a combined efficiency (``normalised trigger performance'' $\hat\epsilon$), defined as the geometric mean 
$$\hat\epsilon = (\epsilon_{Etac} \cdot \epsilon_{Dch} \cdot \epsilon_{Lamc})^{1/3}$$
of the efficiencies of the channels $Etac$, $Dch$ and $Lamc$.

\begin{table}[bp!]
    \centering
    \caption{The list and the description of the NNs investigated for the PANDA Software Trigger in this note.}
    \scriptsize
    \begin{tabular}{ r | l }
        \toprule
        Abbreviation & Description \\
        \midrule
        DNN & Dense NN \\
        DNNRes & Dense NN with residual blocks \\
        CNN & Convolutional NN \\
        CNNRes & CNN with residual blocks \\
        CNNResBN & CNN with Bottleneck residual blocks \\
        LSTM1D & Long-Short-Term Memory with 1D input \\
        LSTM2D & Long-Short-Term Memory with 2D input \\
        \bottomrule
    \end{tabular}\label{tab.NNs}
\end{table}

The absolute value of $\hat\epsilon$ is the primary measure how well a network type performs.
The standard deviation of the individual efficiencies from a series of training runs with varying random initialisation of the internal weights is considered as a measure for the training stability.
Furthermore the networks are investigated for their robustness under changing the network size, in terms of number of layers or blocks, which may be required in a future scenario for more complex triggering setups.

In Figure~\ref{Fig.numBlocks4Compare}, we present exemplarily four network types (DNN, CNN, CNNRes and LSTM1D) and their individual efficiencies for just the channel $Dch$. For each network size ten networks have 
been trained and evaluated, providing a median and standard deviation from the individual results.
The straight forward approach would be a DNN (Figure~\ref{Fig.numBlocks4Compare}, top left), however, the triggering efficiency and stability for training is inferior as it is visible from the large spread of the results.
For a NN with increasing depth, in general significant degradation of performance is expected because of shattered gradients\ \cite{shatteringGradient}.
This can be most clearly seen in case of the CNN with decreasing performance using a
larger number of layers (Figure~\ref{Fig.numBlocks4Compare}, top right).
Using a CNN with residual blocks (Figure~\ref{Fig.numBlocks4Compare}, bottom left) mitigates this behaviour. 
The LSTM model also looks stable but shows a worse generalisation than CNNRes, with regard to the different physics channels (Figure~\ref{Fig.numBlocks4Compare}, bottom right)

For the selection of the best architecture the combined efficiencies $\hat\epsilon$, shown in Figure~\ref{Fig.NNcomparison}, are compared. Here all considered network types are presented, 
which contains the information of a total of $\approx$ 2200 network training and evaluation runs, three channels times ten networks per marker.
We identify the CNNRes type network (orange) as optimal for the given purpose based on its performance in triggering efficiency as well as training stability and robustness concerning varying size.

\subsection{Chosen Network Architecture}
The final choice to evaluate the performance of a neural network approach for the PANDA Software Trigger is the CNNRes, a Convolutional Neural Network with four residual blocks. It is ``deep'' enough to ensure that the NN has enough capacity, while performing relatively stable when adding more blocks, which may be considered within a production triggering environment, e.g.\ featuring many more trigger lines.

\begin{figure}[tp!]
    \centering
    \includegraphics[width=0.65\columnwidth]{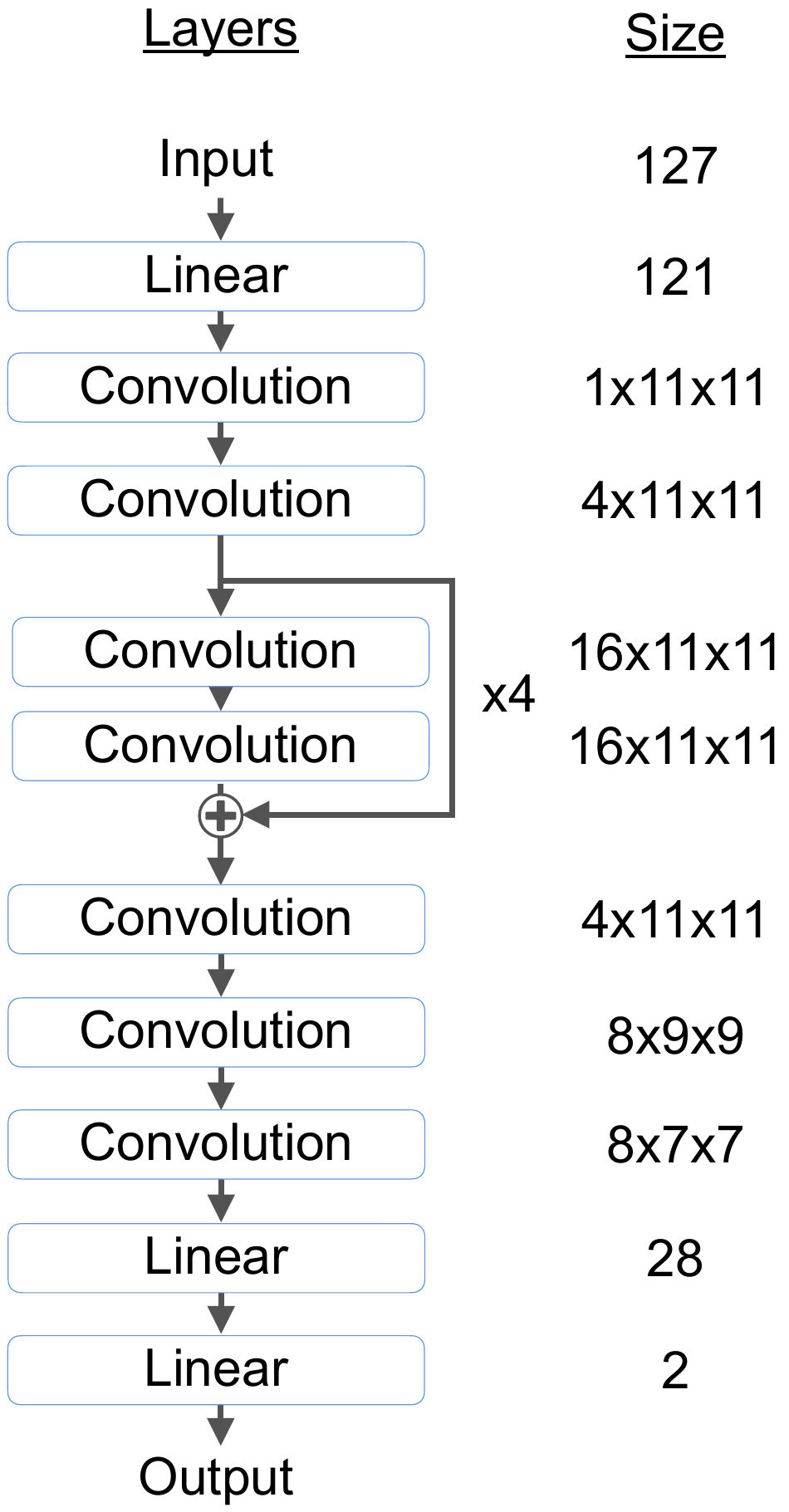}
    \caption{Schematic view of the chosen network architecture.}\label{Fig.NetworkScheme}
\end{figure}

In Figure~\ref{Fig.NetworkScheme}, the details of the network are presented.
In the first stage the 127 input channels are mapped to a 11$\times$11 matrix with a fully connected 121 large linear layer in between. 
Here, the network has the possibility to sort any linear combination of observables next to each other, which will enhance the performance of the image recognition type architecture to follow.
Then two convolutional layers extend the dimensionality to 16$\times$11$\times$11, allowing for a detailed feature extraction performed by four residual blocks~-~pairs of convolutional layers with a residual path.
Three convolutional and two linear layers then reduce the dimensions and perform the classification for the output.
Between each layer batch normalisation\ \cite{BatchNorm} is being performed.

\section{Results and discussion}
\subsection{Individual trigger performance}
Measuring the performance of the neural network approach is done by the individual trigger efficiency determined per channel with that corresponding single trigger line active.
The cuts on the network output are tuned to achieve a background suppression of 1/1000 in total, with an equal fraction of background contribution by each trigger channel for a certain centre-of-mass energy. 
For example at the energy of $\sqrt{s} = 4.5\,\mathrm{GeV}$ with nine of the ten channels being energetically accessible, the targeted background suppression per channel is 1/9000.
Another commonly established measure of the network capabilities are the ROC curves and, more condensed, the corresponding AUCs.
Both, the individual triggering efficiencies and the network AUCs are summarised in Table~\ref{tab.EfficinciesResult}.

\begin{table*}[t]
    \centering
    \caption{Collection of individual trigger efficiency results for a target 1/1000 background suppression, the AUC values of the ROC curves as well as the simultaneous trigger efficiency gains for all accessible channels.}
    \scriptsize
    \begin{tabular}{l | r r r r r r r r r r r}
        \toprule
        eff[\%]  & $ee$      & $Phi$     & $Etac$ & $J2e$     & $J2mu$    & $D0$      & $Dch$     & $Ds$      & $Lam$     & $Lamc$\\
      \bottomrule
      \multicolumn{10}{l}{} \\
      \multicolumn{10}{l}{\bf Individual trigger efficiencies [\%]} \\
      \toprule
        \multicolumn{10}{l}{Conventional trigger} \\
        \midrule
        2.4 GeV &  94.79  &  38.09  &   -   &   -   &   -   &   -   &  -  &   -   &  21.50  &   - \\
        3.8 GeV &  88.80  &  54.64  &   9.71  &  91.34  &  92.52  &  39.50  &12.50  &   -   &  40.25  &   - \\
        4.5 GeV &  98.88  &  56.24  &  11.92  &  85.85  &  84.62  &  31.42  &10.19  &   6.38  &  49.21  &   - \\
        5.5 GeV &  95.94  &  61.47  &  17.87  &  76.31  &  75.09  &  36.79  &17.71  &  11.87  &  56.27  &   6.59 \\
    \toprule
        \multicolumn{10}{l}{NN trigger} \\
        \midrule
        2.4 GeV &   99.98  &   63.67  &     -  &     -  &     -  &     -  &     -  &     -  &   29.14  &     - \\
        3.8 GeV &   99.98  &   78.34  &   22.45  &   99.53  &   99.75  &   39.54  &   23.89  &     -  &   56.67  &     -\\
        4.5 GeV &   99.99  &   78.71  &   21.77  &   98.46  &   99.40  &   54.77  &   24.77  &   15.78  &   63.81  &     -\\
        5.5 GeV &   99.99  &   80.94  &   28.53  &   96.44  &   98.35  &   62.74  &   35.43  &   22.74  &   70.55  &   11.85\\
        \toprule
        \multicolumn{10}{l}{NN trigger with additional event shape observables} \\
        \midrule
        2.4 GeV &  100.00  &   73.72  &     -  &     -  &     -  &     -  &     -  &     -  &   30.33  &     - \\
        3.8 GeV &  100.00  &   85.05  &   26.56  &   99.76  &   99.88  &   69.30  &   30.48  &     -  &   68.52  &     -  \\
        4.5 GeV &  100.00  &   84.13  &   25.21  &   99.21  &   99.64  &   58.01  &   27.32  &   18.92  &   73.47  &     -  \\
        5.5 GeV &  100.00  &   84.95  &   38.10  &   98.86  &   98.90  &   65.02  &   40.16  &   27.26  &   78.19  &   14.75  \\
    \bottomrule
    \multicolumn{10}{l}{} \\
    \multicolumn{10}{l}{\bf Individual NN performance} \\
    \toprule
        \multicolumn{10}{l}{AUC values for the NN trigger} \\
   \midrule
    2.4 GeV	&	1.000	&	0.948	&	-	&	-	&	-	&	-	&	-	&	-	&	0.978	&	-	\\
    3.8 GeV	&	1.000	&	0.987	&	0.869	&	0.999	&	0.999	&	0.960	&	0.854	&	-	&	0.987	&	-	\\
    4.5 GeV	&	1.000	&	0.988	&	0.912	&	0.998	&	0.999	&	0.951	&	0.866	&	0.865	&	0.990	&	-	\\
    5.5 GeV	&	1.000	&	0.986	&	0.929	&	0.996	&	0.998	&	0.936	&	0.890	&	0.852	&	0.992	&	0.821	\\
    \toprule
        \multicolumn{10}{l}{AUC values for the NN trigger with the additional event shape observables } \\
   \midrule
    2.4 GeV	&	1.000	&	0.959	&	-	&	-	&	-	&	-	&	-	&	-	&	0.977	&	-	\\
    3.8 GeV	&	1.000	&	0.988	&	0.887	&	0.999	&	1.000	&	0.963	&	0.838	&	-	&	0.990	&	-	\\
    4.5 GeV	&	1.000	&	0.990	&	0.913	&	0.998	&	0.999	&	0.956	&	0.877	&	0.890	&	0.994	&	-	\\
    5.5 GeV	&	1.000	&	0.989	&	0.941	&	0.999	&	0.998	&	0.955	&	0.900	&	0.873	&	0.994	&	0.834	\\
   \bottomrule
       \multicolumn{10}{l}{} \\
       \multicolumn{10}{l}{\bf Cross-tagging efficiency gains as effect of simultaneous triggering [\%]} \\
       \toprule
        \multicolumn{10}{l}{Conventional trigger} \\
        \midrule
2.4 GeV &  0.01  &  0.43  &  -  &  -  &  -  &  -  &  -  &  -  &  1.96  &  -    \\ 
3.8 GeV &  22.74  &  4.52  &  21.93  &  0.83  &  1.45  &  4.12  &  5.36  &  -  &  3.74  &  -    \\ 
4.5 GeV &  25.53  &  5.68  &  48.65  &  3.68  &  1.00  &  6.94  &  31.82  &  18.51  &  4.12  &  -    \\ 
5.5 GeV &  19.68  &  19.37  &  93.05  &  5.01  &  3.53  &  12.05  &  50.72  &  40.75  &  2.18  &  15.50    \\ 
        \toprule
        \multicolumn{10}{l}{NN trigger} \\
        \midrule
2.4 GeV &   0.00  &  0.00  &  -  &  -  &  -  &  -  &  -  &  -  &  1.26  &  -    \\ 
3.8 GeV &  26.39  &  4.75  &  12.17  &  2.28  &  0.47  &  9.36  &  5.02  &  -  &  3.57  &  -    \\ 
4.5 GeV &  34.92  &  6.51  &  42.59  &  3.42  &  1.75  &  6.12  &  18.13  &  16.22  &  4.58  &  -    \\ 
5.5 GeV &  26.37  &  15.80  &  132.98  &  6.22  &  5.11  &  9.08  &  27.23  &  30.85  &  4.71  &  10.99    \\ 
        \toprule
        \multicolumn{10}{l}{NN trigger with additional event shape observables} \\
        \midrule
2.4 GeV &  0.01  &   0.00  &  -  &  -  &  -  &  -  &  -  &  -  &  0.98  &  -    \\ 
3.8 GeV &  26.93  &  6.72  &  14.57  &  3.92  &  1.23  &  4.13  &  6.48  &  -  &  4.10  &  -    \\ 
4.5 GeV &  35.89  &  7.22  &  61.43  &  4.71  &  3.22  &  6.54  &  22.28  &  16.78  &  4.08  &  -    \\ 
5.5 GeV &  29.18  &  21.76  &  118.00  &  12.93  &  8.09  &  10.37  &  31.27  &  32.09  &  4.15  &  11.46    \\ 
        \bottomrule
    \end{tabular}\label{tab.EfficinciesResult}
\end{table*}

\begin{figure*}[p]
    \centering
    \includegraphics[width=2\columnwidth]{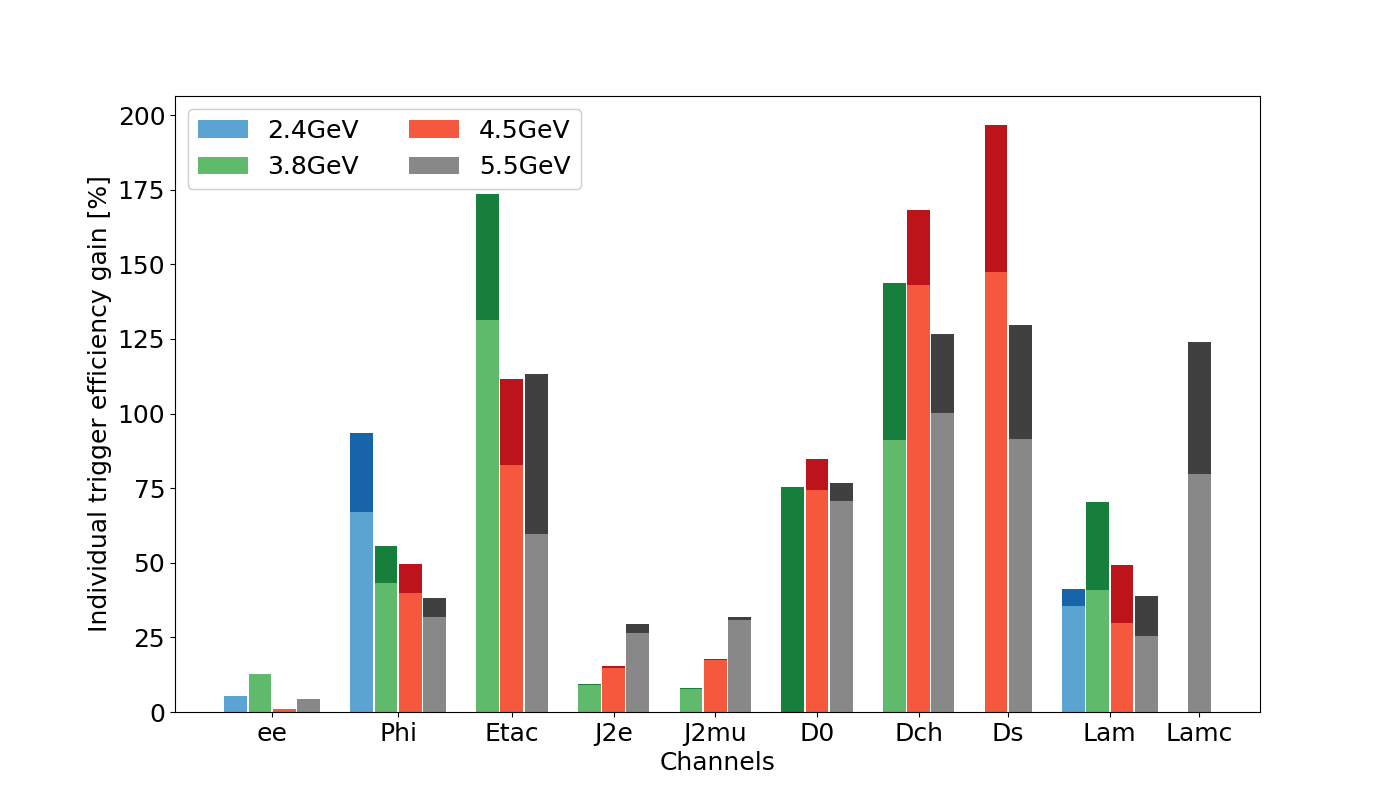}
    \caption{Individual triggering efficiency gains of the neural networks compared to the conventional approach, based on the same set of input observables (light colours), and on the
    extended set of observables (dark colours).}\label{Fig.EffGainBar}
    \centering
    \includegraphics[width=2\columnwidth]{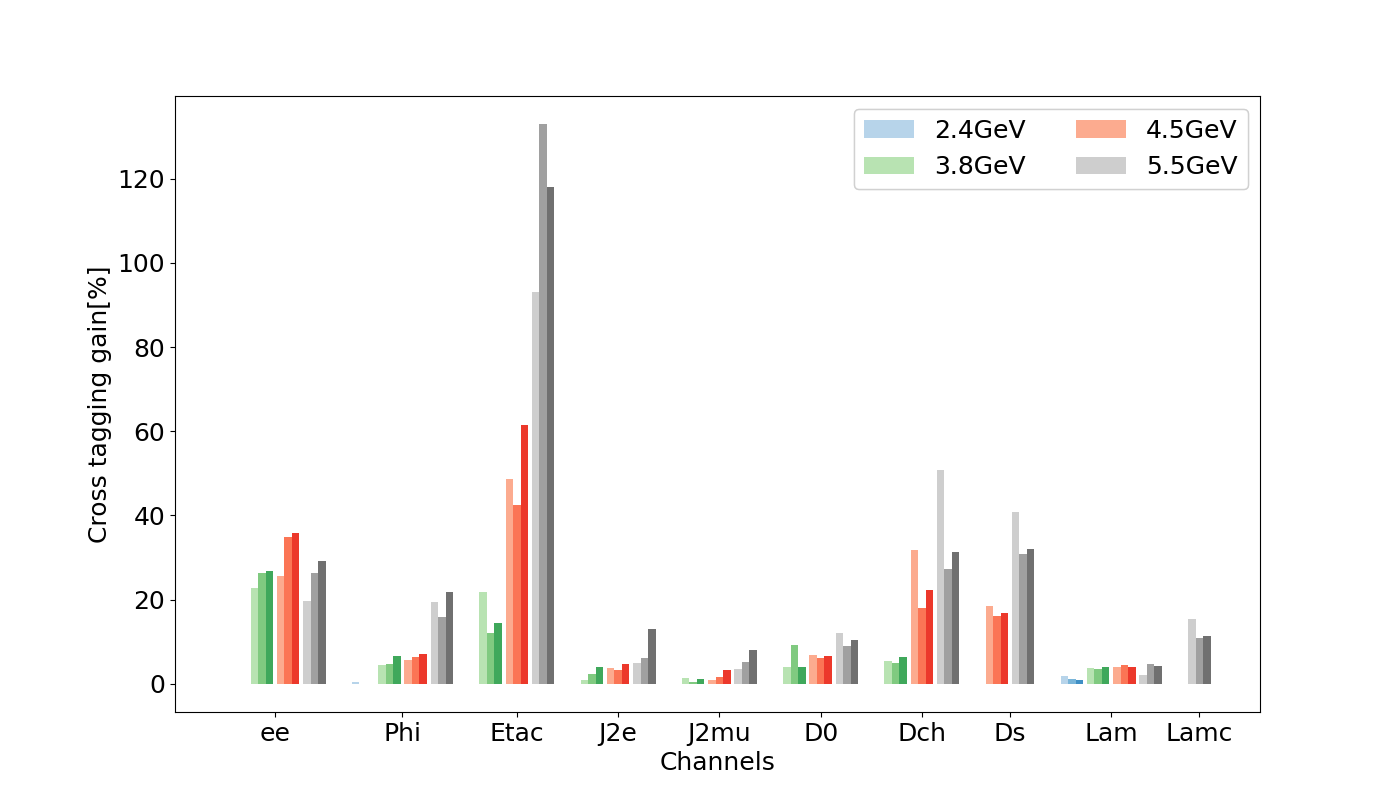}
    \caption{Cross-tagging efficiency gain as effect of simultaneous triggering of multiple reactions, for the conventional benchmark 
    approach (light colours), the NN approach with the same set of observables (medium colours), and the
    NN approach with the extended set of observables (dark colours).}\label{Fig.EffBarAbs}
\end{figure*}


In all cases the triggering with the aid of the neural networks improves the efficiencies compared to the cut-and-count approach. 
This improvement is of course expected, mainly because possible correlations between observables are exploited by the new algorithm.
Figure~\ref{Fig.EffGainBar} shows the relative efficiency 
gain, where the bars in light colours represent the increase based on the same set of input variables, the dark coloured bars the performance
gain using the extended set of variables. It reaches almost up to 200\% for some of the channels,
corresponding to about a factor three in performance. 
The actual benefit strongly depends on the channel in question. For example the  channel $ee$ has already a very good triggering efficiency from the cut-and-count approach, leaving no room for improvements.
In the case of the two $J/\psi$ channels $J2e$ and $J2mu$, the triggering efficiency is dropping with increasing beam momentum (see Table~\ref{tab.EfficinciesResult}), which could be almost completely recovered by the new approach.
For the open charm and $Etac$ and $Lamc$ channels the triggering efficiency has about doubled.

Adding more observables, describing the overall event topology, leads in many cases to a substantial gain in triggering efficiency.
Hence, in a future software trigger setup it would be desirable to provide these event shape observables, even if computationally expensive in an online environment.
Based on the AUC values all networks show good (0.8 - 0.9) to excellent ($> 0.9$) classification performance.

\subsection{Simultaneous trigger performance}
In a realistic setup, all active trigger lines have the ability to trigger events simultaneously. Therefore it can happen, that a signal event of a
certain type missed by the dedicated trigger line is accidentally tagged by another one.
This cross tagging by simultaneous triggering has the potential to further improve the triggering efficiency.
Nevertheless, at the same time the background level will not exceed the requirements.

Figure~\ref{Fig.EffBarAbs} quantifies this effect in the current scenario, showing the increase in the triggering efficiencies, also presented in Table~\ref{tab.EfficinciesResult}, which are calculated as the fraction of 
triggered events from events that passed the reconstruction in the trigger line designed for the channel.
The light, medium and dark coloured bars correspond to the conventional cut-and-count benchmark algorithm, the neural network approach and the neural network approach based on the extended variable set, respectively.
In some cases this adds a substantial increase in triggering efficiency. For example the channel $Etac$ at 5.5~GeV gains a factor of two in total triggering efficiency by cross tagging in this particular setup.
It is important to note that these ``accidental'' gains are highly dependent on the actual composition of trigger lines, the centre-of-mass energy, and the actual trained networks.

\begin{figure*}[tp!]
    \centering
   \includegraphics[width=1.\columnwidth]{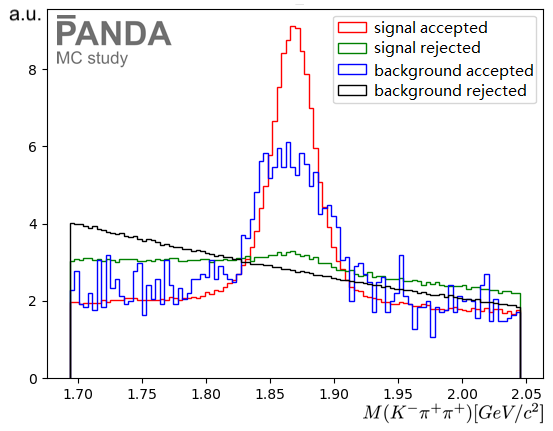}
   \includegraphics[width=1.\columnwidth]{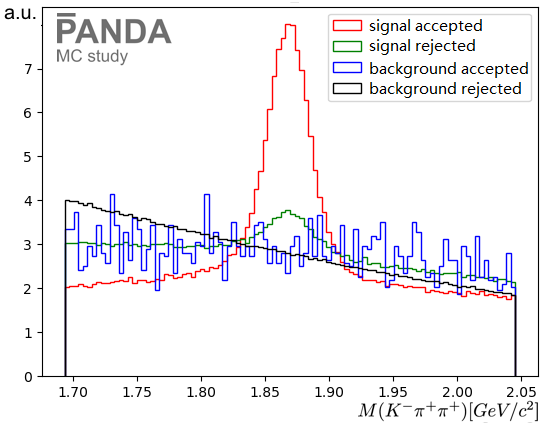}
    \caption{Candidate mass distributions of accepted and rejected signal (red and green, respectively) and background (blue and black, respectively) for the channel $Dch$  at 4.5~GeV/$c$. Left, all kinematic input
     observables, right without the energy of the candidate and the invariant mass of the recoiling system. The histograms are normalised to the same integral.}\label{Fig.DchBgPeak}
\end{figure*}

\subsection{Computing performance}
The bulk of training processes goes into finding the optimal solution for each network architecture, which is time consuming and requires the intensive usage of GPU and CPU resources.
A dedicated server with one Intel\R\ Xeon\R\ W-2135 CPU with 3.70~GHz and one NVIDIA\R\ 3090 GPU is used,  equivalent in performance to 179 cores of the AMD\R\ EPYC\R\ 7551 32-Core Processors on the computing cluster at GSI.
Processing times are in the order of two days for a single trigger line on 12 cores of the computing cluster.
Since the training can be performed in parallel for the trigger lines this is quite promising.
The model inference is able to run in the order of 1.4~M candidates per second on the GPU server with six trigger lines in parallel. The main bottleneck is the CPU based data reading and preparation. 

\subsection{Data quality and choice of observables}
The input features and observables have varying impact on the trigger decision. 
Some features are correlated, some are not relevant for the classification problem at hand, which differs between the channels and even between the data sets of the same channel at different beam energies.
The ranking of the observable importance can guide the observable choice and thus may reduce the required computations in the online processing before.

Furthermore, it is important to maintain a certain quality of the data during the selection process, which in turn means that the underlying physics constrains the choice of observables.

The $Dch$ channel ($D^+\rightarrow K^-\pi^+\pi^+$) for example has a significant enhancement in the invariant mass distribution of background events around the signal peak position (blue histogram in Figure~\ref
{Fig.DchBgPeak}, left) 
when using both the energy of the $D^+$ candidate and the invariant mass of the recoil system at the same time as input for the NN. This effect originates from the specific kinematics of
the two body reaction $\bar{p}p\rightarrow D^+D^-$ considered here.  
However, it is the goal to trigger $D^+$ particles from many different reactions inclusively.
Dropping these observables from the NN input removes the undesirable peaking background (Figure~\ref{Fig.DchBgPeak}, right) at the cost of some signal efficiency. We observed now a smooth distribution (blue histogram) in the 
signal region similar to the rejected background (black histogram).

\begin{figure*}[tp!]
    \centering
    \includegraphics[width=1\columnwidth,trim=20 0 50 35, clip]{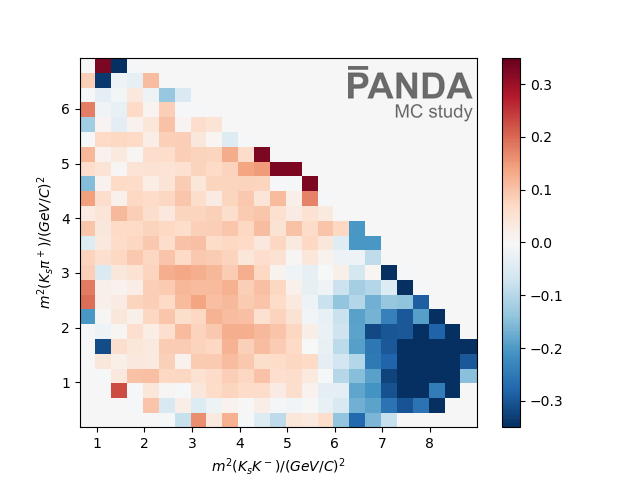}
    \includegraphics[width=1\columnwidth,trim=20 0 50 35, clip]{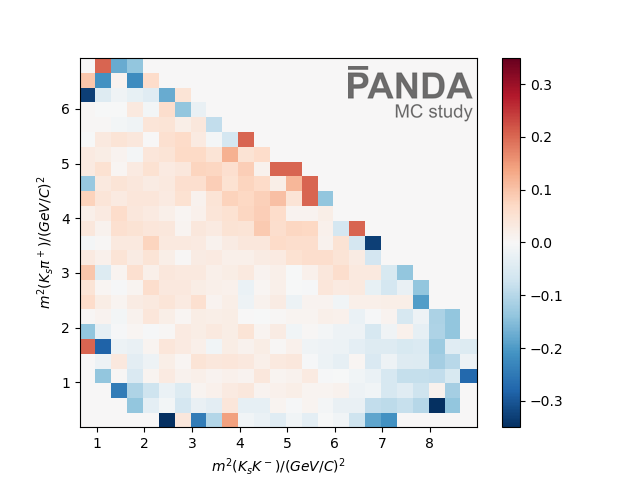}
    \caption{Trigger efficiency deviation from the mean value in Dalitz coordinates for the channel $Etac$ at 3.8~GeV/$c$, without (left) and with (right) the additional observables.}\label{Fig.Dalitz}
\end{figure*}

When analysing complex decay patterns and mixtures of resonances, it is important to cover the phase space without steep drops or holes in the triggering efficiency distributions introduced by the triggering algorithm. 
Complex fitting algorithms, such as a partial-wave analysis \cite{PrimerPWA}, benefit most from a flat triggering efficiency distribution if possible, but they require at least a smooth dependency on any kinematic observable.

For example a three-body-decay, such as the channel $\eta_c\rightarrow K_sK^-\pi^+$, could be studied in the Dalitz plot\ \cite{Dalitz} representation.
We find that the relative efficiency of the neural network trigger introduces a cut-off in one corner of the Dalitz plot (Figure~\ref{Fig.Dalitz}, left) with the observables that are used in the cut-and-count approach. It 
could be that the network identified this corner to be particularly occupied with background.
Introducing the event shape observables (Table~\ref{tab.Observables}) as additional input, the training is able to produce a significantly more flat triggering efficiency (Figure~\ref{Fig.Dalitz}, right), being beneficial for 
later Dalitz plot analysis.

\section{Summary and outlook}
Based on a set of ten channels with typical event topologies in the reach of PANDA physics investigated at four anti-proton beam energies, 
it is demonstrated, that PANDA will greatly benefit from a neural network supported software trigger system.

As one result the use of binary classification networks outperforms an approach with multi-class classification. 
As an additional advantage, it will make the setup much easier to extend the system with a larger set of simultaneous trigger lines and improves the computational scalability.
From seven network architectures, the convolutional network with four residual blocks showed the best results for the three channels with the lowest triggering efficiency, while producing stable results under different 
training attempts and size changes.

In all cases the network approach performs better than the cut-and-count method, which further improves when adding more observables. 
In the comparison the triggering efficiencies show gains of up to 200\% and all networks show good up to excellent performance.

Data quality is an important topic which deserves to be studied in more detail. 
It would be desirable to give a feedback about e.g.\ the triggering efficiency flatness in the Dalitz plot coordinates to the neural networks during the training process.
This also holds for the background flatness in critical distributions, such as the candidate invariant mass in order to avoid peaking structures.

For the study presented here, the background suppression requirement is set equally among the physics channels. 
However, this approach neglects the differences in reconstruction and varying background levels in the different channels.
Finding an approach to achieve the best background suppression requirements for each trigger line, while maintaining the desired overall background reduction, is a complex challenge. 
Enhancing the triggering efficiency of one channel will reduce the triggering efficiency of another channel, which has to be carefully balanced in the future experiment.


\begin{acknowledgements}
We acknowledge the internal review work of the accompanying analysis memo by A.~Belias, M.~Papenbrock and T.~Stockmanns, and we thank all members of the PANDA Collaboration.
This work has been supported by the Bundesministerium f\"ur Bildung und Forschung (BMBF), Germany;
the GSI Helmholtzzentrum f\"r Schwerionenforschung GmbH, Germany; the Helmholtz Forschungsakademie Hessen f\"ur FAIR (HFHF), Germany; the Helmholtz-Institut Mainz, Germany;
the Institute of Modern Physics, Chinese Academy of Sciences, China. And in particular, P.~Jiang has received financial support via the Helmholtz -- OCPC Postdoc Program and the Office of China Postdoc Council, Ministry of Human Resources and Social Security, China.
\end{acknowledgements}



%

\end{document}